\documentclass[12pt,a4paper]{article}

\usepackage{amsmath}
\usepackage{amssymb}
\usepackage{graphicx}
\usepackage[left=2cm,top=2cm,right=2cm,bottom=2cm,nohead]{geometry}
\usepackage{color}
\usepackage{epstopdf}
\definecolor{Red}{rgb}{1.0, 0.0, 0.0}

\begin{document}

\title{Topological Qubit Design and Leakage}
\author{R.~Ainsworth\,$^1$ and J.\,K.~Slingerland\,$^{1,2}$\\
\textit{\footnotesize
{}$^1$ Department of Mathematical Physics, National University of Ireland}  \\
\textit{\footnotesize Maynooth, Co.~Kildare, Ireland}\\
\textit{\footnotesize {}$^2$ Dublin Institute for Advanced Studies, School for Theoretical Physics}\\
\textit{\footnotesize 10 Burlington Rd, Dublin, Ireland} \\
\footnotesize E-mail: \texttt{robert.ainsworth@nuim.ie}, \texttt{joost@thphys.nuim.ie}
}
\date{\small February 21, 2011}
\maketitle

\abstract{%
We examine how best to design qubits for use in topological quantum computation. 
These qubits are topological Hilbert spaces associated with small groups of anyons. Operations are performed on these by exchanging the anyons. One might argue that, in order to have as many simple single qubit operations as possible, the number of anyons per group should be maximized. However, we show that there is a maximal number of particles per qubit, namely $4$, and more generally a maximal number of particles for qudits of dimension $d$.  
We also look at the possibility of having topological qubits for which one can perform two-qubit gates without leakage into non-computational states. It turns out that the requirement that all two-qubit gates are leakage free is very restrictive and this property can only be realized for two-qubit systems related to Ising-like anyon models, which do not allow for universal quantum computation by braiding. 
Our results follow directly from the representation theory of braid groups which means they are valid for all anyon models. We also make some remarks on generalizations to other exchange groups. 
}

\section{Introduction}

Topological quantum computation \cite{Kitaev03,Freedman98} has been with us for well over a decade now, and despite the great difficulty in producing a first experimental realization of the idea, may represent our best hope for the future of quantum computing, due to its inherent fault tolerance and potential for scalability. 

Concrete proposals for topological quantum computation center around two dimensional systems with excitations which have nontrivial behavior under exchanges, non-Abelian anyons. Great theoretical and experimental effort has been made to understand the various systems in which anyons are thought or hoped to occur and a good deal is also known about the types of computations that could be done if the anyons were given to us and we were allowed to exchange them at will in any way we like. Perhaps most importantly, it is known for a large class of anyon models that they provide for universal quantum computation by exchanges \cite{Freedman02a,Freedman02b}. A sort of standard paradigm for the construction of a topological quantum computer has also emerged and versions of this occur in papers such as \cite{hzbs1,hzbs2} and \cite{fklw1}. The idea is to use the topological quantum numbers of small groups of anyons as qubits and to perform operations on these qubits by exchanging the anyons, both within the groups that form the qubits and, for multi-qubit gates, also between groups. We will go into a little more detail about this in section~\ref{sec:paradigm}. The importance of such a paradigm is that it allows to make direct contact with the circuit model of quantum computation and it enables algorithmic questions to be tackled independently of the details of experimental implementation, at least initially. 

The aim of this paper is to study some constraints on the TQC paradigm which exist for any anyon model. These arise when one asks some very basic questions, such as \emph{`What is the optimal number of anyons with which to form a qubit~?'} and \emph{`To what extent can the problem of leakage be avoided~?'}. Leakage is the issue that, because the full Hilbert space of a multi-anyon system does not have a tensor product structure, a subspace of this Hilbert space must be chosen if one wants to deal with qubits, and on exchanging the anyons, some of the topological amplitude may `leak' into the non-computational states. While these questions must certainly have been considered by the authors of earlier papers, we feel they have so far not really been brought to the forefront. Nevertheless, some interesting general results can be obtained which are valid for all types of anyons. For example, qubits with groups of $5$ or more anyons are not possible, unless one allows for leakage in single-qubit gates. Similar constraints apply to \emph{qudits}, the $d$-dimensional analogues of qubits. It also turns out that very few topological 2-qubit systems allow for totally leakage free braiding of the anyons and those which do will not allow for universal braiding. Again, similar results can be obtained for some systems with higher dimensional qudits. 

The structure of the paper is as follows. We start with a brief overview of the standard paradigm for TQC and the leakage problem which is inherent to it. Then we focus on single qubits and qudits and show that, if exchanges within the groups of anyons forming the qudits are to be leakage free, the number of particles that form the qudits is limited. Our arguments actually show that similar constraints on the number of particles per qudit should hold for any computational scheme based on exchanging localized objects, for example for closed string excitations in three dimensions. We then consider two-qubit and two-qudit gates and we show that very few situations exist in which all exchanges of the anyons and hence all conceivable two-qudit gates, are leakage free, at least for low-dimensional qudits. moreover, the exceptions we have identified do not allow for universal quantum computation by braiding. Finally we make some general remarks on leakage free gates in systems which do have leaking gates, and conclude. 

\subsection{Standard TQC scheme and leakage}
\label{sec:paradigm}

\begin{figure}[htb]
\begin{minipage}[h!]{0.4\textwidth}
\centering
\hspace*{0.15\textwidth}
\includegraphics[scale=0.5]{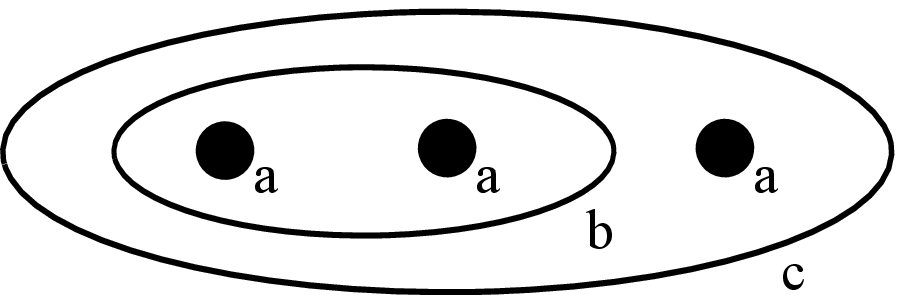}\\
(a)
\end{minipage}
\hspace*{-0.05\textwidth}
\begin{minipage}[h!]{0.4\textwidth}\hspace*{0.1\textwidth}
\centering
\includegraphics[scale=0.4]{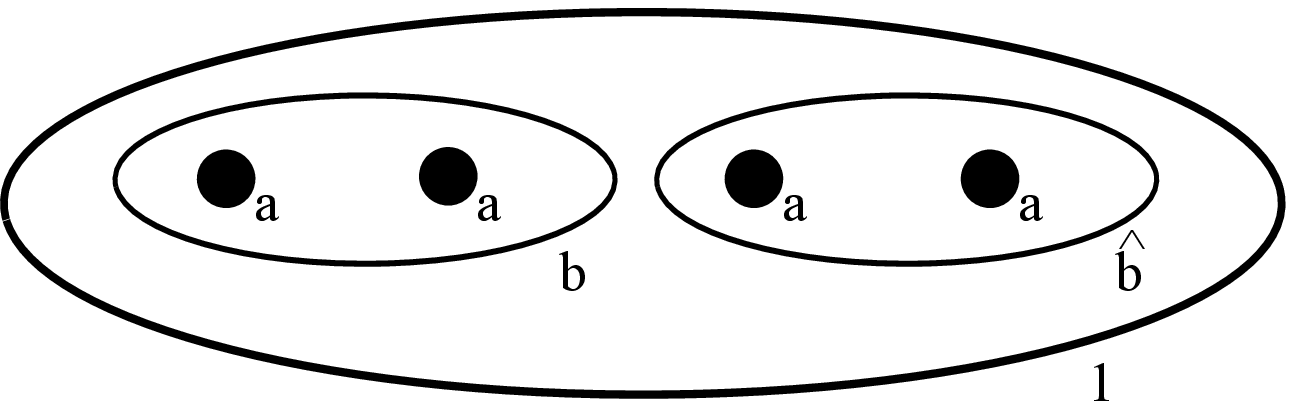}\\
(b)
\end{minipage}\\
\hspace*{2cm}\parbox{0.8\textwidth}{\caption{\label{fig:qubit} \footnotesize \textbf{(a)} A qudit composed of three anyons of topological charge $a$. The information in the qudit is stored using the topological charge $b$ of the first two. All three together have a fixed topological charge $c$. \textbf{(b)} A qudit composed of four anyons of topological charge $a$. The information is stored using the topological charge $b$ of the first two. The total topological charge is trivial ($1$) and the last two automatically have the topological charge $\hat{b}$, which can fuse to $1$ with $b$.}}
\end{figure}

In the `standard model' of a topological computer, our system is composed of a collection of anyons, described by an anyon model or topological field theory (see e.g.~\cite{Preskill-lectures,Kitaev06a,Bonderson_PhD} for reasonably `physicist friendly' introductions to anyon models). The Hilbert space of the entire system is the fusion space of the anyons, which has a basis of states labeled by the different ways in which all the anyons can fuse together. This space typically does not have a natural tensor product structure, so we introduce this by hand; we associate qudits with small groups of anyons. Usually one takes these groups to consist of three or four anyons, as indicated in figure~\ref{fig:qubit}. 

The overall topological charge of each such group of anyons is usually fixed, and is conserved by exchange processes involving only the anyons in a single qudit. For 4-anyon qudits, one may choose this topological charge to be the trivial, or vacuum charge, which is convenient since it means the qubit as a whole can be exchanged without any nontrivial effect on the system's state.  The quantum number which stores the information in the qudit is the topological charge of the left-most two anyons in the qudit.
The computational Hilbert space is taken to be the tensor product of these qudit spaces and the topological quantum computer is then effectively described by the quantum circuit model, with the important addition that there is a natural set of gates generating all possible quantum operations, namely the elementary exchanges of the anyons in the qudits.  

The computational Hilbert space, however, is usually only a subspace of the full topological Hilbert space of the system. While braiding anyons from the same qudit does not change the overall topological charge of the qudit, braiding anyons from different qudits around each other may change the overall topological charge of the qudits and this will result in some of the information \lq leaking' out of the computational Hilbert space. Moreover, if such leakage has happened at some stage in a computation, the amplitude in the non-computational subspace may couple back into the computational subspace later on.
  
In the all models of this sort proposed so far (see e.g.~\cite{hzbs1,hzbs2,fklw1}) leakage is unavoidable and it is of great importance to find gates which leak as little as possible and in general, to reduce the amount of leakage in a system to a minimum (see~\cite{xuwan1} for work specifically focused on this). Clearly, in systems where exchange processes allow for an approximation of arbitrary unitaries, it is always possible to avoid leakage to arbitrary accuracy, but avoiding leakage is nevertheless a nuisance which can cause severe overhead. In this paper we will examine under which circumstances one may have models where all 1-qudit and/or all multi-qudit braiding gates are entirely leakage free. Such models would obviously be extremely desirable as they would make computations simpler and more accurate. However, it turns out that excluding leakage is a very severe restriction on the allowable models.

\section{Design of a Single Qudit}
\label{sec:qudit_design}

We would like to find the optimal number of anyons from which to construct a single qudit. Obviously, what is optimal depends on the circumstances, for example, if it is experimentally hard to control multiple anyons, then the optimal number may be the minimal number of anyons which actually can provide a qudit - this would be two anyons, with the qudit of information stored in the fusion channel of the two anyons. However, using 2-anyon qudits has a clear disadvantage, which is that in a system with 2-anyon qudits, no nontrivial gates can be performed by braiding anyons within a single qudit. As a result, leakage enters even at the level of single qudit gates -- in fact it may be difficult to perform many exact single qubit gates, because they would have to be built from special braidings of anyons that are part of different qudits. Qudits with $3$ anyons solve this problem and can even allow for a computationally universal set of single-qudit operations which are implemented by exchanges within the qudits and hence leakage free. Qudits with $4$ anyons have the additional advantage that they can have trivial total topological charge. When this is the case, the qudits can be moved around the quantum computer without disrupting the computation, since exchanges with the trivial charge have no effect on the state of the system. If we set aside issues of controlling large numbers of anyons, it appears that the optimal number of anyons per qudit may be the largest possible number (we will soon show that there is a maximum, in a useful sense). By increasing the number of anyons in each qubit we increase the number of elementary exchange operations which are possible within each qubit. This has two advantages, firstly with more elementary operations it may be easier to achieve a desired unitary operation and secondly it could reduce the need to use two-qudit gates which would introduce leakage errors. We will therefore be interested in finding the maximal number of anyons per qudit.

\begin{figure}[htb]
\centering
\includegraphics[scale=0.4]{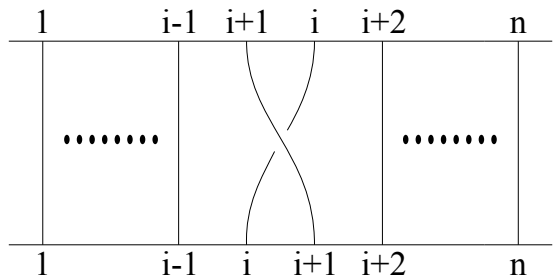}
\hspace*{1cm}\parbox{0.8\textwidth}{\caption{\footnotesize \label{fig:exchange}
The exchange of the $i^{th}$ and $i+1^{th}$ anyon, i.e. the braid $\tau_{i}$. We may think of this braid diagram as a spacetime diagram of the exchange process. Applying consecutive exchanges then corresponds to stacking such diagrams, or braid group multiplication.}}
\end{figure}

To study qudits containing $n$ anyons, we will use the braid group, $B_{n}$ which governs the exchanges of the $n$ anyons inside the qudit. $B_{n}$ has $n-1$ generators, $\tau_{1},\ldots,\tau_{n-1}$, where $\tau_{i}$ exchanges particle $i$ with its neighbor, particle $i+1$ (see figure~\ref{fig:exchange}). These generators obey the following relations:
\begin{eqnarray}
\tau_{i}\tau_{j}&=&\tau_{j}\tau_{i} ~~\lbrace \mathrm{provided~} |i-j|\geq 2\rbrace
\label{eq:commrel}\\
\tau_{i}\tau_{i+1}\tau_{i}&=&\tau_{i+1}\tau_{i}\tau_{i+1}
\label{eq:ybrel}
\end{eqnarray}

$B_{n}$ will act on the $d$-dimensional Hilbert space of the qubit by a unitary representation. In principle, we may choose the $d$-dimensional qudit Hilbert space as a subspace of some higher dimensional representation of $B_{n}$, but we will not be interested in that possibility here, because in such cases, either $B_{n}$ mixes the qudit Hilbert space with the other states in the representation, which means that leakage occurs even for braidings within the qudit, or alternatively, the qudit's Hilbert space is preserved under the action of $B_{n}$, but then we may as well just consider it by itself as a $d$-dimensional representation of $B_{n}$. 

For any $n$, there are $1$-dimensional unitary representations of $B_n$ labeled by an exchange angle $\theta$ and explicitly given by $\tau_{j}\mapsto e^{i\theta}$ (for all $j$). Clearly this means that $d$-dimensional representations of $B_{n}$ can be constructed for any $n$ by taking direct sums of such one-dimensional representations; in such representations all $\tau_{i}$ can be diagonalized simultaneously. Such representations are not very useful for TQC. Therefore, given the dimension $d$ of our qudit, we will look for the maximal number of particles $n$ for which we can have a $d$-dimensional representation of $B_n$ which is not of this reducible form. 

\subsection{Qubits}
\subsubsection{No leakage free qubits with more than $\mathbf{4}$ anyons}
\label{sec:qubit34}

As stated in the introduction, previous work on qubit design has concentrated on working with models where each qubit contains three or four anyons. We will give a simple argument that shows that there are in fact no qubits with $n>4$ anyons which have a nontrivial representation of $B_n$.

To study $2$-dimensional representations of $B_{n}$, we first choose a basis such that $\tau_{1}$ is a diagonal matrix. Relation~(\ref{eq:commrel}) then tells us that any generator which is not adjacent to $\tau_{1}$ must commute with it and relation~(\ref{eq:ybrel}) tells us that if neighboring generators commute then they are equal. Moreover, it is not hard to see that all the generators $\tau_{i}$ are conjugate to each other and this implies that if any pair of neighboring generators $\tau_{i},\tau_{i+1}$ commute, all neighboring pairs commute and in fact as a result, if any neighboring generators commute, then all generators are represented by the same matrix, i.e.~by the same diagonal matrix as $\tau_1$.  

If the two eigenvalues of $\tau_{1}$ are equal then $\tau_{1}$ is a multiple of the identity matrix and hence commutes with all other generators, giving us a diagonal representation. We therefore look at the case where the eigenvalues of $\tau_1$ are not equal. Now we note the fact  that, if a diagonal $d\times d$-matrix has $d$ distinct eigenvalues, any other matrix which commutes with it  must also be diagonal. If $N\ge 5$, we have generators, $\tau_{1},\ldots,\tau_{3},\tau_{4},\ldots$ and then~(\ref{eq:commrel}) shows that both $\tau_{3}$ and $\tau_{4}$ commute with $\tau_{1}$. As a result, for $d=2$, both $\tau_{3}$ and $\tau_{4}$ must be diagonal. But then $\tau_{3}$ and $\tau_{4}$ commute with each other which, by~(\ref{eq:ybrel}), gives us a diagonal representation.

It is clear therefore that to avoid our representation of the braid group becoming diagonal (i.e.~a sum of  two $1$-dimensional representations), we must construct our qubits out of no more than four particles. If we do decide to have qubits with more than four anyons then we have to choose between having qubits without interesting single qubit operations from in-qubit braiding, or introducing leakage for single qubit operations. Neither of these alternatives seem desirable so we should ensure that $N \leq 4$. This shows that the models which have so far been examined do, in fact, use the optimal numbers of anyons per qubit.

\subsubsection{Braiding and Universality for Qubits}

Since qubits with $3$ or $4$ anyons are the only ones which allow interesting internal braiding and no leakage, let us give explicit formulae for the elementary exchange matrices as they act in an arbitrary unitary two dimensional representation of $B_3$ or $B_4$ (these will also be useful for future reference). In any representation $B_{n}\rightarrow U(d)$, we may divide all representation matrices by their determinant to obtain a representation of $B_n$ into $SU(d)$. The determinant is an overall phase and hence irrelevant to TQC, but we may always reobtain all $U(d)$ representations from the $SU(d)$ representations by multiplying these by $1$-dimensional unitary representations (given by $\tau_{j}\mapsto e^{i\phi}$ for some $\phi\in\mathbb{R}$). In the case of a qubit, we can thus work with $SU(2)$ matrices.  Let us start with the case of $B_3$. We will call our arbitrary representation $\eta:B_{3}\rightarrow SU(2)$. It is convenient to choose a basis in which $\eta(\tau_1)$ is diagonal. We will assume its eigenvalues $a$ and $\bar{a}$ are distinct, since otherwise the whole representation will be diagonal. Using up the remaining freedom in our choice of basis, we can ensure that the off-diagonal elements of $\eta(\tau_2)$ are real and the off diagonal element in the first row of the matrix is positive. We can then directly use the defining relations of $B_3$ to obtain the possible forms for the generators. We find that
\begin{equation}
\label{eq:qubitrep}
\eta(\tau_{1})=
\left(\begin{matrix}
a & 0 \\
0 & \bar{a} \end{matrix}\right)
~~~~
\eta(\tau_{2})=\left(\begin{matrix}
\frac{1}{a-a^{3}} & b \\
-b & \frac{1}{\bar{a}-\bar{a}^{3}} \end{matrix}\right),
\end{equation}
where $a=e^{i\theta}$ for some $\theta\in \mathbb{R}$, and $b=\sqrt{1-\frac{1}{2-a^{2}-\bar{a}^{2}}}$.
Since $b$ was chosen to be real, and therefore $b^2>0$, there are restrictions on the value of $a$. Taking the determinant of $\tau_{2}$ gives $\frac{1}{2-a^{2}-\bar{a}^{2}}+b^{2}=1$ and as a result $\frac{1}{2-a^{2}-\bar{a}^{2}} \leq 1$. Substituting $a=e^{i\theta}$, we get the restriction $\frac{-\pi}{6} \leq \theta \leq \frac{\pi}{6}$.

If the qubit is composed of four anyons we get one extra generator, $\tau_{3}$. Since the generators are all conjugate, they must have the same eigenvalues and since $\eta(\tau_3)$ commutes with $\eta(\tau_1)$ it must be diagonal. Hence, it has two possible values: $\eta(\tau_{3})=\eta(\tau_{1})$ or $\eta(\tau_{3})=\bar{\tau_{1}}$. If $\eta(\tau_{3})=\eta(\tau_{1})$ this poses no extra restrictions on $a$ or $b$. However if $\eta(\tau_{3})=\bar{\tau_{1}}$ then one finds from $\tau_2\tau_3\tau_2=\tau_3\tau_2\tau_3$ that $a$ must be a primitive $8^{th}$ root of unity, that is $a=\pm e^{\frac{\pm i \pi}{4}}$.

Any representation of the braid group which comes from an anyon model with finitely many distinct topological charges must have eigenvalues for the exchanges $\tau_i$ which are roots of unity at some finite order. This is guaranteed by Vafa's theorem (see \cite{Vafa88} and also the appendix of \cite{Kitaev06a}). In other words, not all values of $a$ which are allowed by the braid group relations are likely to occur. It is of interest to identify the Jones representations of $B_3$, which are related to $SU(2)$ Chern-Simons theory, among the full set of representations given here. 
Explicit matrices $\rho_{r}(\tau_1)$ and $\rho_{r}(\tau_2)$ for these representations are given in~\cite{Freedman02b}, in the following form,
\begin{equation}
\rho_{r}(\tau_{1})=\left(\begin{matrix}
q & 0 \\
0 & -1 \end{matrix}\right)
~~~~~
\rho_{r}(\tau_{2})=\left(\begin{matrix}
\frac{-1}{q+1} & \sqrt{[3]_{q}} \\
\sqrt{[3]_{q}} & \frac{q^{2}}{q+1} \end{matrix}\right),
\end{equation}
where $[x]_{q}=\frac{q^{x/2}-q^{-x/2}}{q^{1/2}-q^{-1/2}}$ and $q=e^{\pm\frac{2\pi i}{r}}$, with $r\in\mathbb{N}$, $r\ge3$.

We can see from this that if we multiply our representation, $\eta$, by a factor of $-a$ and perform a coordinate transformation such that the off-diagonal terms pick up a factor of $i$, then the two representations will be equal, provided $q=-a^{2}$.
This means that $q=-e^{2i\theta}=e^{i(2\theta-\pi)}$. But $\frac{-\pi}{6} \leq \theta \leq \frac{\pi}{6}$, so taking the two extreme values of $\theta$ we get $q=e^{\pm \frac{2i\pi}{3}}$. 

We can now make some strong statements about the universality of the $B_3$ and $B_4$ representations we have found. Freedman, Larsen and Wang's universality results in \cite{Freedman02b} reduce in this simple case to the statement that the images of the representations $\rho_{r}$ are dense in an $SU(2)$ subgroup of $U(2)$ whenever $r \geq 5$, $r \neq 6,10$. Hence our corresponding representations are dense in $SU(2)$ for the corresponding values of $a$. 

Recent results of Kuperberg~\cite{kuperberg1} actually settle the question of universality for any other roots of unity and even for arbitrary eigenvalues as well. His corollary 1.2 and theorem 1.4 state, for the representations of $B_3$ and $B_4$ considered here, that their images are all dense in $SU(2)$, unless $q=-a^2=e^{i\phi}$ with $|\phi|=\pi-\frac{2\pi}{n}$, where $\phi$ is an angle and $n\in\mathbb{Z}$, $n\ge 3$, or $q$ is a root of unity of order $10$. Note that the cases $n=3$, $n=4$ and $n=6$ are the only cases in which $\pi-\frac{2\pi}{n}$ is of the form $\frac{2\pi}{r}$ for an integer $r$, with also $r\in\{3,4,6\}$ so that the exceptional cases with $\phi=\pm\frac{2\pi}{r}$ are the same as those found by Freedman, Larsen and Wang. 

\subsection{Qudits with $\mathbf{d>2}$}
\label{sec:qudits}

In section~\ref{sec:qubit34}, we showed that a qubits should involve $n<5$ anyons in order to allow for leakage free and non-Abelian single qubit operations. We now argue that there is a similar maximum number of anyons for qudits of any dimension $d$. Our arguments will not depend strongly on the detailed structure on the braid group, beyond features which are common to many similar exchange groups, such as the motion group for ring-shaped excitations in three dimensions and other motion groups, as defined by Dahm~\cite{Dahm62} (see also~\cite{goldsmith}). Specifically, we will use the following properties of the braid group representation which acts on the qudit:
\begin{enumerate}
\item Generators which do not involve the same objects (i.e. particles, strands) commute.
\item The group is represented unitarily.
\item The generators in our favored set are conjugate to each other and in fact any adjacent pair of generators is conjugate to any other adjacent pair.
\end{enumerate}
The first of these properties is connected with the basic physical principle that spatially separated operators commute and the second comes from the unitarity of time evolution. The availability of a set of conjugate generators is a bit more special. However, one may expect this whenever, as in the case of the braid group, all generators perform the same type of action and the ordering of the objects being exchanged is just a matter of convention. If there are different types of generators (such as exchanges of distinguishable types of particles), then this will no longer fully hold. 

\subsubsection{Existence of an upper limit on $\mathbf{n}$ for qudits}

A representation of $B_n$ will become completely abelian if any generator commutes with a neighboring generator. {}From property 3 above, it then follows that all generators will commute with those adjacent to them and hence with all other generators (using property 1). Let us consider a $d$-dimensional representation $\eta$ of $B_{n}$. Since the generators of $B_{n}$ are conjugate, their representation matrices $\eta(\tau_{i})$ all have the same eigenvalues $(\lambda_1,\ldots,\lambda_{d})$. We now choose a basis in which all the representation matrices of the odd numbered generators, $\tau_{1},\tau_{3,\ldots}$ are simultaneously diagonal. For $n\ge 5$, no two of the diagonal matrices $\eta(\tau_{i})$ with $i$ odd can be the same, unless $\eta$ is Abelian. The reason for this is that, when $n\ge 5$, for every pair of odd generators $\tau_{i}$, $\tau_{j}$ (with $i,j$ odd) there is an even generator which is adjacent to one of the pair, let's say $\tau_i$, but not adjacent to the other, say $\tau_j$. If $\eta(\tau_{i})=\eta(\tau_{j})$, then the even generator adjacent to $\tau_{i}$ will commute with $\eta(\tau_{i})$, since it necessarily commutes with $\eta(\tau_{j})$ and $\eta(\tau_{i})=\eta_(\tau_{j})$. But then two neighboring generators commute and as we argued before, the whole representation becomes Abelian. Hence a non-Abelian $d$-dimensional representation of $B_n$ must yield a different arrangement of the eigenvalues $(\lambda_1,\ldots,\lambda_d)$ in every one of the representation matrices of the odd generators $\tau_{1},\tau_{3},\ldots$ when these are diagonalized. Since there are only a finite number of possible distinct arrangements of the eigenvalues (at most $d!$ if they are all distinct), this places an upper limit on the number of odd generators, and hence on the number of anyons one may have at this value of $d$ without making the representation completely Abelian. 

\subsubsection{Qutrits}

The previous paragraph shows that there is an upper limit on the number of anyons per qudit for any $d$, but it obviously overestimates this limit. It is interesting to look in a bit more detail at the case of \emph{qutrits}, i.e.~$d=3$. As previously stated, we can diagonalise all the odd numbered generators. If all eigenvalues are equal we will always have a trivial representation and if all are different we will get the same result as we obtained for qubits in section~\ref{sec:qubit34}, namely we must have $n\le 4$. There is then only one case left for us to look at, the case with two distinct eigenvalues, one with multiplicity $2$ and one with multiplicity $1$. There are only three patterns in which such a selection of eigenvalues can be arranged in a diagonal matrix therefore we know we must have no more than three odd numbered generators. If we now look at the even numbered generators we see that there can be at most three, $\tau_{2}$, $\tau_{4}$ and $\tau_{6}$. The form of the representation matrices of the even numbered generators is restricted, as they must commute with all spatially separated odd numbered generators, and hence must preserve their eigenspaces. 

We see then that, for $n=7$, $\tau_{2}$ and $\tau_{4}$ will preserve the same eigenspaces as $\tau_{5}$ and $\tau_{1}$, respectively, but $\tau_{6}$ will commute with both $\tau_{1}$ and $\tau_{3}$ and so must preserve the same eigenspaces as both of these generators. But $\tau_{1}$ and $\tau_{3}$ must have different patterns of eigenvalues. Let us denote our basis vectors $\beta_1$, $\beta_2$ and $\beta_3$, then by a choice of ordering, we can make sure that $\beta_1$ and $\beta_2$ form the two dimensional eigenspace of $\tau_1$ and $\beta_2$ and $\beta_3$ form the two dimensional eigenspace of $\tau_3$. Since $\tau_6$ must conserve both eigenspaces, it cannot map $\beta_2$ into $\beta_1$ or $\beta_3$ and in fact we see that $\tau_{6}$ must be diagonal. As a result, $\tau_{6}$  commutes with $\tau_{5}$ and so we have two neighboring generators commuting and our representation of $B_{7}$ is abelian.

Moving to the $n=6$ case, we eliminate the problem of having an even numbered generator which commutes with two odd numbered generators but nevertheless, this case too proves to be Abelian. We now have two even generators, $\tau_{2}$ and $\tau_{4}$, which both commute with only one odd numbered generator, but we also have to take into consideration the fact that all even generators are non-neighboring and so they must commute among themselves. However, the form of the matrices representing $\tau_{2}$ and $\tau_{4}$ is restricted in different ways, because they preserve different eigenspaces of different odd numbered generators. By a suitable ordering of the basis, we can make sure that $\tau_2$ and $\tau_4$ map to matrices of the folowing form,
\[
\tau_2\mapsto
\left(\begin{matrix}
M^{(1)}_{11} & M^{(1)}_{12}&0 \\
M^{(1)}_{21} & M^{(1)}_{22}&0 \\
0&0&\lambda
\end{matrix}\right)
~~~~~
\tau_4\mapsto
\left(\begin{matrix}
\lambda&0&0\\
0&M^{(2)}_{11} & M^{(2)}_{12} \\
0&M^{(2)}_{21} & M^{(2)}_{22} \\
\end{matrix}\right),
\]
where $M^{(1)}$ and $M^{(2)}$ are unitary $2\times 2$-matrices and $\lambda$ is the eigenvalue of the $\tau_i$ which has multiplicity $2$. 
The only way two such matrices may commute is if they are both diagonal and in that case the representation becomes abelian. In conclusion, we see that for $d=3$ we must have $n\le 5$.
Notice that we have never explicitly used the braid relation $\tau_{i}\tau_{i+1}\tau_{i}=\tau_{i+1}\tau_{i}\tau_{i+1}$ up to now, though of course it has been present in the background, in the sense that it guarantees that the elementary exchanges are all conjugate to each other. Moving to higher dimensions $d$ makes this kind of argument more difficult as the basis vectors can be split into groups such that all of the groups contain more than one vector, which makes it more difficult to determine the precise nature of the restrictions which come from the fact that even generators commute. 

\subsubsection{Some general definitions and results}
\label{sec:generaldefs}

Clearly, the actual maximum number $n$ for which we may have a non-Abelian, leakage free qudit of dimension $d$ involving $n$ anyons, or equivalently, a non-Abelian $d$-dimensional representation of $B_{n}$, depends strongly on the number and multiplicity of the eigenvalues of the generators of the braid group in the representation. The number of distinct eigenvalues of an elementary exchange is usually equal to the number of distinct topological charges that may be produced in the fusion of two of the elementary anyons in the system. Hence we should not expect this to be very large, typically.  The actual values of the eigenvalues are connected with the topological spins, or conformal weights of the various fusion products of the elementary anyons and are important observables in many physical implementations of anyonic systems. Therefore, it would seem to make sense to refine our search for the qudit with the largest number of anyons. We can define various versions of the `maximal number of anyons' as follows. 
\begin{itemize}
\item
$N(d)$ is the largest $n$ for which $B_{n}$ has a non-Abelian representation of dimension $d$. 
\item
$N(d,p)$ is the largest $n$ for which $B_{n}$ has a non-Abelian representation of dimension $d$, such that the elementary exchanges have $p$ distinct eigenvalues. 
\item
$N(d,\bar{m})$ is the largest $n$ for which $B_n$ has a non-Abelian $d$-dimensional representation such that the exchanges have eigenvalues with multiplicities $m_1,\ldots,m_p\,$ given by the partition $\bar{m}$ of $d$. For example if $\bar{m}=(2,2,1)$ then $d=5$ and the representation is required to have three distinct eigenvalues, two of them with multiplicity $2$ and one with multiplicity $1$
\item
$N(d,\bar{m},\bar{\lambda})$ is the largest $n$ for which there is a non-Abelian $d$-dimensional representation of $B_{n}$ such that the exchanges have eigenvalues $\bar{\lambda}=(\lambda_{1},\ldots,\lambda_{p})$ with multiplicities $\bar{m}=(m_{1},\ldots,m_{p})$.
\end{itemize}
Any attempt at a full determination of all these numbers is well beyond the scope of this paper, but the arguments we have given already provide upper bounds for all these numbers (though not very good ones). Our results for qubits and qutrits may be summarized as $N(2)=N(2,2)=4$, $N(3)=N(3,2)=N(3,(2,1))=5$ and $N(3,3)=4$. It is easy to see that our argument for qubits generalizes to all cases where the number of eigenvalues equals $d$ and we have $N(d,d)=4$. Many interesting representations with two and three distinct exchange eigenvalues (corresponding to two or three fusion channels for the elementary anyons) are obtained from the theory of Hecke algebras and BMW-algebras and it would be interesting to apply these to find more of the $N(d,\bar{m},\bar{\lambda})$. It is actually not difficult to see (see also below) that a unitary representation with two eigenvalues and dimension $d$ exists for $n=d+2$, so $N(d,2)\le d+2$ (and we will see below that this implies that $N(d,2)=N(d)=d+2$) but the question becomes more interesting if the multiplicities or values of the eigenvalues are given. 

An important general result for $N(d)$ has been proved by Formanek in ref.~\cite{formanek1}. This paper shows that 
\[N(d)=d+2,\] 
which is exactly the result we have found above in the $d=2,3$ cases. We nevertheless gave our own proofs for these special cases, because we wanted to make it obvious that these results are based on the simple physical properties of exchange groups highlighted at the start of section~\ref{sec:qudits}, namely the fact that the operations are conjugate to each other and non-neighboring generators commute. We therefore expect that the result found for $N(d)$ can be generalized to apply to a range of exchange groups beyond $B_{n}$.

All the irreducible representations of the braid group $B_{n}$ which are of dimension $d \leq n$, have in fact been classified relatively recently in ref.~\cite{flsv1}. There are eleven cases, $(A)$ through $(K)$, listed in that paper. Cases $(A)$ and $(B)$, shown below, apply to any number of particles, while the remaining cases are special cases and we shall mention only the ones which are necessary for our calculations.
\begin{itemize}
\item[(A)] A representation of Burau type, either:\\
$\chi(y) \otimes \beta_{n}(z): B_{n} \rightarrow GL_{n-1}(\mathbb{C})$, where $1+z+...+z^{n-1} \neq 0$ or\\
$\chi(y) \otimes \widehat{\beta_{n}}(z): B_{n} \rightarrow GL_{n-2}(\mathbb{C})$, where $1+z+...+z^{n-1}=0$.
\item[(B)] 
A representation of standard type: $\chi(y) \otimes \gamma_{n}(z): B_{n} \rightarrow GL_{n}(\mathbb{C})$, where $z \neq 1$.
\end{itemize}

\noindent In both cases $\chi$ is a character of $B_{n}$ (i.e. a one dimensional representation). Also $\beta_{n}(z)$ denotes the reduced Burau representation of $B_{n}$ with parameter $z$, while $\widehat{\beta_n}$ is the nontrivial composition factor of $\beta_{n}(z)$ which exists when $z$ is an $n^{th}$ root of unity. Explicit formulae for $\beta_{n}$ and $\widehat{\beta_n}$ and also for the standard representation $\gamma_{n}$ can be found in~\cite{flsv1}.

The special cases all occur for $3 < n < 9$. 
We will mention the cases which are relevant to qutrits. These necessarily occur for $2\le n\le 5$. The case $n=2$ is uninteresting, because necessarily Abelian. For $n=3$, we have case $(B)$ above.
For $n=4$, we have $\beta_{4}$ from $(A)$, as well as the special case (D). The representation in case $(D)$ is written $\epsilon_{4}(z): B_{4} \rightarrow GL_{3}(\mathbb{C})$ and is given in ref.~\cite{flsv1}. Finally for $n=5$, we use the nontrivial composition factor $\widehat{\beta_{5}}(z)$ from $(A)$.

Ref.~\cite{flsv1} deals with representations into $GL_{d}(\mathbb{C})$ so one may be concerned about unitary. However, we find that, in each of the above cases, taking $\chi(y)$ unitary and restricting the values of the parameter $z$ to roots of unity yields a unitary (or at least unitarizable) representation.

\section{Two Qudit Gates and Leakage}
\label{sec:2qudits}

Even if we assume that no leakage occurs in single qubit gates, it is still likely to plague multi-qubit gates. On the other hand, if we are able to construct a universal set of leakage free $2$-qubit gates, then we can do universal quantum computation without any leakage, since any multi-qubit gates can be generated from $1$-qubit and $2$-qubit gates. A first simple question one may ask is whether systems of two qubits (or more general qudits) exist in which all anyon exchange processes are leakage free. This means in effect that the computational Hilbert space for such a $2$-qudit system is closed under the action of the braid group for the anyons involved in the qudits. In other words, the computational Hilbert space itself carries a representation of the full two qubit braid group. Let us assume that the two qubits involve $n$ anyons, with the individual qudits consisting of $n_1$ and $n_2$ anyons (where $n_1+n_2=n$). Any representation $\rho$ of $B_n$ which provides us with leakage free 2-qubit braiding then has some very special properties. First of all, if the individual qudits have Hilbert space dimensions $d_1$ and $d_2$ then the $2$-qudit Hilbert space has dimension $d=d_1 d_2$ and we can think of it as the tensor product of the qudits' Hilbert spaces. Now, for the elementary exchanges $\tau_{1},\ldots,\tau_{n_1-1}$ which involve only the anyons in the first qudit, we can write the representation matrices as follows
\begin{equation}
\label{eq:leftqudit}
\rho(\tau_{i})=\rho_{1}(\tau_{i}) \otimes \mathbb{I}_{d_{2}}~~~~~(1\le i\le n_{1}-1),
\end{equation}
where $\rho_{1}$ is the $d_1$-dimensional representation of $B_{n_1}$ on the Hilbert space of the first qudit and $\mathbb{I}_{d_{2}}$ is the $d_{2}$-dimensional identity matrix. 
Similarly, the matrices for the elementary exchanges $\tau_{n_1+1},\ldots,\tau_{n-1}$ which involve only the anyons in the second qudit, can be written as follows
\begin{equation}
\label{eq:rightqudit}
\rho(\tau_{i})=\mathbb{I}_{d_{1}} \otimes \rho_{2}(\tau_{i})~~~~~(n_1+1\le i\le n-1),
\end{equation}
where $\rho_{2}$ is the $d_2$-dimensional representation of $B_{n_2}$ on the Hilbert space of the second qudit.
Once the single qudit representations $\rho_1$ and $\rho_2$ are fixed, 
the representation matrix for the only remaining braid generator, $\tau_{n_1}$, is subject to the constraint that it must solve all the braid relations (\ref{eq:commrel}) and (\ref{eq:ybrel}). Whenever a solution to these constraints can be found, we have a completely leakage free two-qubit system. However, we will see that most of the time, the constraints cannot be satisfied, so that for almost all types of qubits, it is unavoidable that leakage will appear for at least some of the possible exchange processes.

\subsection{Two Qubits}
\label{sec:2qubits}

We showed in section 2.1 that a leakage free, non-Abelian qubit can be made of either three or four anyons, therefore there are in principle three different types of $2$-qubit system, firstly with two qubits each composed of three anyons ($2 \times 3$-qubit), secondly with two qubits each composed of four anyons ($2 \times 4$-qubit) and lastly with one $3$-qubit and one $4$-qubit. However, for the $2\times 4$-qubit system we would need to have a four dimensional representation of $B_{8}$ and for the case with a 4-qubit and a 3-qubit we would need a four-dimensional representation of $B_{7}$ to obtain totally leakage free braiding. As we discussed in section~\ref{sec:generaldefs}, it was shown in~\cite{formanek1} that no $d$-dimensional representation of $B_n$ can exist with $d<n-2$ and therefore both of these cases cannot be realized without leakage. 

This leaves us with the system of two $3$-anyon qubits to consider. We need to construct a 4-dimensional representation, $\rho$, of the braid matrices of the group $B_{6}$. For this two-qubit system to be leakage free we require it to decompose as a tensor product of two two-dimensional representations. We have five elementary exchanges, $\tau_{1},\ldots, \tau_{5}$. Of these, $\tau_{1}, \tau_{2}$ and  $\tau_{4},\tau_{5}$ will take the forms given in equations (\ref{eq:leftqudit}) and (\ref{eq:rightqudit}) respectively, and by a convenient choice of basis, we can make sure that the single qubit representations $\rho_1$ and $\rho_2$ which occur in those equations are both given by the standard form $\eta$ which we introduced in equation~(\ref{eq:qubitrep}). This yields 
\begin{center}
$\rho(\tau_{1})= \left(\begin{matrix}
a & 0 & 0 & 0 \\
0 & a & 0 & 0 \\
0 & 0 & \bar{a} & 0 \\
0 & 0 & 0 & \bar{a}
\end{matrix}\right)$ \hspace{75pt} $\rho(\tau_{2})= \left(\begin{matrix}
\frac{1}{a-a^{3}} & 0 & c & 0 \\
0 & \frac{1}{a-a^{3}} & 0 & c \\
-c & 0 & \frac{1}{\bar{a}-\bar{a}^{3}} & 0 \\
0 & -c & 0 & \frac{1}{\bar{a}-\bar{a}^{3}}
\end{matrix}\right)$

$\rho(\tau_{4})= \left(\begin{matrix}
\frac{1}{f-f^{3}} & e & 0 & 0 \\
-e & \frac{1}{\bar{f}-\bar{f}^{3}} & 0 & 0 \\
0 & 0 & \frac{1}{f-f^{3}} & e \\
0 & 0 & -e & \frac{1}{\bar{f}-\bar{f}^{3}}
\end{matrix}\right)$ \hspace{75pt} $\rho(\tau_{5})= \left(\begin{matrix}
f & 0 & 0 & 0 \\
0 & \bar{f} & 0 & 0 \\
0 & 0 & f & 0 \\
0 & 0 & 0 & \bar{f}
\end{matrix}\right)$
\end{center}
All the generators are conjugate and have the same eigenvalues, so $f=a$ or $\bar{a}$, and hence also $e=c=\sqrt{1-\frac{1}{2-a^{2}-\bar{a}^{2}}}$. We can now fix $\rho(\tau_3)$ by imposing the braid group relations. Since $\tau_{3}$ is unitary and commutes with $\tau_{1}$ and $\tau_{5}$, we see that it must be diagonal. It will also have eigenvalues $a$ and $\bar{a}$, like the other generators, but it cannot have the same arrangement of eigenvalues as $\tau_{1}$ or $\tau_{5}$ as this will mean one of the adjacent generators will commute with it and we will get an abelian group (cf.~section~\ref{sec:qudits}). Hence,
\begin{center}
$ \rho(\tau_{3})= \left(\begin{matrix}
x & 0 & 0 & 0 \\
0 & \bar{x} & 0 & 0 \\
0 & 0 & \bar{x} & 0 \\
0 & 0 & 0 & x
\end{matrix}\right) $
$\lbrace$where $x=a$ or $x=\bar{a} \rbrace$
\end{center}
The relation $\tau_{2}\tau_{3}\tau_{2}=\tau_{3}\tau_{2}\tau_{3}$ now yields the following two equations,
\begin{center}
(8) $\displaystyle \bar{c}\left[\frac{x}{a-a^{3}}-\frac{\bar{x}}{\bar{a}-\bar{a}^{3}}\right]=\bar{c}|x|^{2}$\\
(9) $\displaystyle \bar{c}\left[\frac{\bar{x}}{a-a^{3}}-\frac{x}{\bar{a}-\bar{a}^{3}}\right]=\bar{c}|x|^{2}$.
\end{center}
We can equate the left hand sides of eqns (8) and (9) and by some simple manipulations we get a restriction on the eigenvalue $a$:
\begin{center}\vspace{-10pt}
$a^{2}=-\bar{a}^{2}$
\end{center}\vspace{-10pt}
But $a=e^{i\theta}$ so we get $e^{2i\theta}=-e^{-2i\theta}$, or $cos(2\theta)=-cos(2\theta)$. It follows that $cos(2\theta)=0$ which gives $\theta = \pm \frac{\pi}{4}$ or $\theta=\pm \frac{3\pi}{4}$. In short, $a$ must be a primitive $8^{th}$ root of unity. One now checks easily that with these values of $a$, we also satisfy the relation $\tau_3\tau_4\tau_3=\tau_4\tau_3\tau_4$ and so we find that a leakage free representation $\rho$ exists provided the eigenvalues of the elementary exchange matrices are primitive $8^{th}$ roots of unity. Unfortunately, this restriction on the eigenvalues makes the system non-universal for quantum computation (even at the single-qubit level). The representations of $B_6$ we have found in this way are precisely the ones one obtains from anyon models with the fusion rules of the Ising model and in fact, for these anyon models the full $6$-anyon Hilbert space with trivial total topological charge is $4$-dimensional, which explains the absence of leakage. 

\subsection{Qutrits and Qubit-Qutrit combinations}

We can use the method explained at the start of section~\ref{sec:2qudits} also to study whether totally leakage free braiding is possible in systems consisting of two qutrits or systems which have a qubit and a qutrit. All $3$-dimensional representations of $B_{n}$ are known and can be divided into just a few classes, as explained at the end of section~\ref{sec:generaldefs}, with explicit formulae for the representations given in~\cite{flsv1}, so the analysis proceeds in the same way as for two qubit systems. 

Non-Abelian qutrits can be composed of $3$, $4$ or $5$ anyons and so a two qutrit system can have $3+3$ or $3+4$ or $3+5$ or $4+4$ or $4+5$ or $5+5$ anyons. We have checked all cases in the same way as for the $3+3$ anyon system of two qubits in section~\ref{sec:2qubits} and found that no leakage free non-Abelian braid group representations exist, so having at least some braids with leakage is unavoidable in two qutrit systems. 

For a system of one qubit and one qutrit, we are looking for a $B_n$ representation of dimension $d=6$, with necessarily $n\le 8$. Hence we must consider qubit-qutrit systems of $3+3$, $3+4$, $3+5$ or  $4+4$ anyons. Again, we checked all these cases by direct calculation and found no leakage free non-Abelian braid group representations. 

One may in principle go at least a little beyond qutrits by the same method, since for example all four and five dimensional irreducible representations of $B_3$ are given in~\cite{Tuba01}

\subsection{Leakage Free Gates in a System With Leakage}

So far we have shown that for systems consisting of qubits and qutrits, it is not possible to have a situation where all braidings are leakage free, except in an exceptional case with non-universal qubits of Ising type. However, the requirement that all braids are leakage free is needlessly restrictive. Even if leakage occurs for some braids, we could avoid those and do computations using  only those braids which cause no leakage. In fact, to obtain universal leakage free quantum computation by braiding, it is enough to have a single leakage free entangling 2-qubit braiding gate in combination with a universal set of leakage free single qubit gates. Clearly, the braids which do not cause leakage in a certain representation form a subgroup of the braid group, which we may dub the \emph{leakage free subgroup}. It would be of extreme interest to find the leakage free subgroup of the braid group for representations which occur in simple anyon models, and of even greater interest to find the closure of the images of these representations in the corresponding unitary groups. Again, if braiding within qubits is universal and a single leakage free entangling 2-qubit gate exists, then the projective image of the leakage free subgroup should be dense. Unfortunately, no one has so far been able to construct a leakage free entangling $2$-qudit braiding gate in any anyon model which has universality for single qudit operations by braiding, and we will not attempt to change that situation here. Pessimists may be tempted to conjecture at this point that the image of the leakage free subgroup is always either finite or composed of combinations of single qudit gates. 

\section{Summary and Discussion}
We have considered some basic questions concerning the optimal design of qudits in topological quantum computers, notably the question of how many anyons we should use per qudit and the question to what extent leakage can be avoided in single and multi-qudit systems. 
In section~\ref{sec:qudit_design}, we determined that topological qubits can not be composed of more than $4$ anyons, and for general qudits of dimension $d$, there is an upper limit of $d+2$ anyons. We also saw that this upper limit may be lowered if restrictions are placed on the eigenvalues of the elementary exchanges of adjacent anyons. We summarized the properties of qubits with $3$ and $4$ anyons and indicated the full set of braid group representations which could be used for qutrits. In section~\ref{sec:2qudits}, we examined the possibility of creating leakage-free two-qubit gates and found that we can usually not require all braid operations to be leakage-free, and in the exceptional cases where this is possible, we can't have universality. Similar results were obtained for systems which are built from qutrits or from a mixture of qubits and qutrits. 

While our results are only a very small step in what will surely become a great journey toward the optimal framework for topological quantum computation, we hope they highlight some interesting directions which have so far not received very much attention. In particular, it would be very interesting to obtain more information on the various numbers $N(d,p)$, $N(d,p,\bar{m})$ etc.~defined in section~\ref{sec:generaldefs}. Apart from obtaining the actual numbers, one might for example ask in which cases these maxima are obtained by a braid group representation which describes the exchanges of anyons and is hence embedded in a topological quantum field theory. Any progress on the determination of the leakage free group, even for a single anyon model with universal braiding  in single qubits would also be of great importance. 

\subsection*{Acknowledgments}
This work was supported by Science Foundation Ireland Principal Investigator award 08/IN.1/I1961.

\bibliography{leakagebib}
\bibliographystyle{unsrt}

\end{document}